%% file: main.tex
\documentclass[11pt,a4paper]{article}
\usepackage[T1]{fontenc}
\usepackage[utf8]{inputenc}
\usepackage{lmodern}
\usepackage[margin=22mm,headheight=14pt]{geometry}
\usepackage{amsmath,amssymb}
\usepackage{graphicx}
\usepackage{booktabs}
\usepackage{pifont}
\usepackage{xcolor}
\usepackage{microtype}
\usepackage[super,sort&compress]{natbib}
\input{sala-reference-format}

\usepackage{caption}
\usepackage{url}
\usepackage{hyperref}
\hypersetup{colorlinks=true,linkcolor=black,citecolor=black,urlcolor=blue,
  pdftitle={Symmetry-Aware Flow Matching for End-to-end Molecular Crystal Generation},
  pdfsubject={SALA research manuscript}}
\renewcommand{\figurename}{Fig.}
\input{sections/figure_captions}

\title{\vspace{-1.2cm}\bfseries Symmetry-Aware Flow Matching for End-to-end Molecular Crystal Generation}
\author{%
Wendi Cai\textsuperscript{1,\textdagger} and Fanyang Mo\textsuperscript{1,2,3,4,*}\\[0.5em]
\begin{minipage}{0.96\textwidth}
\raggedright\small
\textsuperscript{1}School of AI for Science, Peking University Shenzhen Graduate School, Shenzhen 518055, China\\
\textsuperscript{2}School of Advanced Materials, Peking University Shenzhen Graduate School, Shenzhen 518055, China\\
\textsuperscript{3}Guangdong Provincial Key Laboratory of Nano-Micro Materials Research, Peking University Shenzhen Graduate School, Shenzhen 518055, China\\
\textsuperscript{4}State Key Laboratory of Advanced Waterproof Materials, School of Materials Science and Engineering, Peking University, Beijing 100871, China\\[\baselineskip]
\textsuperscript{\textdagger}Work completed during an internship at ByteDance.\\
\textsuperscript{*}Corresponding author: Fanyang Mo (\href{mailto:fmo@pku.edu.cn}{fmo@pku.edu.cn})
\end{minipage}}
\date{}

\begin{document}
\maketitle
\thispagestyle{plain}

\begin{abstract}
\input{sections/abstract}
\end{abstract}

\input{sections/introduction}
\input{sections/results}
\input{sections/discussion}

\clearpage
\input{sections/methods}
\input{sections/declarations}

\clearpage
\input{sections/extended_data}

\clearpage
\begingroup
\raggedright
\bibliographystyle{sala-numeric}
\bibliography{references}
\endgroup

\end{document}

%% file: sala-reference-format.tex
\setcitestyle{super,comma}
\newcommand{\salaReferenceFirst}{1}
\newcommand{\salaReferenceLast}{9999}
\newcommand{\salaReferenceOffset}{0}

\newcommand{\salaReference}[2]{%
  \ifnum#1<\salaReferenceFirst\relax\else
    \ifnum#1>\salaReferenceLast\relax\else
      \setcounter{NAT@ctr}{\numexpr#1+\salaReferenceOffset-1\relax}%
      #2%
    \fi
  \fi}

%% file: sections/figure_captions.tex
\newcommand{\captionFigOne}{%
\textbf{SALA separates a symmetry-constrained state from full-cell context.}
\textbf{a}, Reduction in independently generated lattice and atomic degrees of freedom; the resulting SALA trajectory remains compatible with the specified space group throughout generation.
\textbf{b}, Setting-specific lattice masks (filled circles, active; open circles, fixed).
\textbf{c}, Paired ASU and full-cell heavy-atom counts on logarithmic axes; median counts increase from 38 in the ASU to 189 in the full cell.
\textbf{d}, SALA data flow.
The Gaussian prior state is sampled only over ASU coordinates and active lattice variables.
The constrained lattice decoder and OrbitExpand construct a symmetry-compatible full cell, represented by three lattice tokens and one token per full-cell heavy atom.
A DiT processes these tokens using chemical, relational and periodic-geometric pair biases; OrbitPullback returns atom predictions to the ASU before the updated cell is reconstructed.
The lower pathway shows training-time augmentation over symmetry-equivalent ASU representatives and closest-periodic-image geometry.
}

\newcommand{\captionFigTwo}{%
\textbf{SALA achieves high hit rates across chemical classes and crystal systems.}
\textbf{a}, Representative crystals illustrating chemical, molecular and packing diversity; translucent overlays show the reference structures, and solid renderings show structures generated by SALA.
\textbf{b}, Three evaluated structural levels: molecular conformation, unit-cell geometry and intermolecular packing.
\textbf{c}, Five corresponding metrics: conformer RMSD, relative volume error, Hencky shape error, whole-supercell packing RMSD and severe-overlap prevalence.
Continuous metrics are reported as means of target-best values obtained by minimizing each metric independently within a candidate pool; overlap prevalence pools candidate-level outcomes.
\textbf{d,e}, Hit rates by crystal system (\textbf{d}) and chemical class (\textbf{e}), where a hit requires at least one clash-free candidate with packing RMSD below \(2\,\text{\AA}\).
SALA generates 50 candidates per target; CLARI uses target-specific pools of 50--3,150 candidates, with details of the allocation protocol provided in the Supplementary Information.
}

\newcommand{\captionFigThree}{%
\textbf{SALA's packing advantage widens with molecular and crystallographic complexity.}
\textbf{a}, Mean target-best packing RMSD stratified by ASU heavy-atom count, full-cell heavy-atom count, molecular-component copies in the cell and symmetry-operation count.
\textbf{b}, Mean paired packing-error difference (CLARI minus SALA) jointly stratified by symmetry-operation count and, from top to bottom, ASU size, full-cell size and component copies.
Positive values favour SALA.
Each of the paired targets contributes its minimum finite packing RMSD and receives equal weight within a bin.
Annotations give differences in \AA\ and target counts; hatched cells contain no targets.
}

\newcommand{\captionFigFour}{%
\textbf{Trajectory-resolved structural organization and attention redistribution in SALA.}
\textbf{a}, Progress of clean-endpoint predictions towards the final output of the same trajectory for unit-cell geometry, conformation and packing, together with intermolecular-contact overlap.
\textbf{b}, Attention enrichment trajectories for pair classes defined by the final structure; the dashed curve denotes atom-to-cell attention.
\textbf{c}, First-to-last enrichment changes for full attention and for bias-only attention recomputed without query--key content logits.
\textbf{d}, Distance--time maps of attention redistribution for intramolecular and intermolecular pairs.
\textbf{e}, Event-aligned bidirectional pair attention and separation around persistent intermolecular-contact formation.
The analysis uses ten samples per target and 24 model evaluations.
Calculation and aggregation procedures are described in Methods, ``Generation dynamics and attention analyses''.
}

\newcommand{\captionFigFive}{%
\textbf{Ablations of SALA's state--context framework and structure-aware design.}
Rows compare full SALA with seven separately trained variants grouped by context, orbit handling and representative augmentation; lattice prior and constrained coordinates; and pair bias and self-conditioning.
Columns show conformer RMSD, relative volume error, Hencky shape error, packing RMSD and samples with severe overlaps; lower values are better.
Each configuration uses the same candidate budget per target.
Continuous metrics are mean target-best values over targets evaluable across all configurations; overlap prevalence pools candidate-level outcomes.
Points denote means or percentages.
}

\newcommand{\captionExtendedDataOne}{%
\textbf{State--context separation across SALA's generative trajectory.}
Given the ASU composition and space group, SALA evolves ASU atomic coordinates and lattice variables as the independent generative state.
At each step, space-group expansion reconstructs the full cell used to predict the next update, without assigning independent states to symmetry-related images.
}

\newcommand{\captionExtendedDataTwo}{%
\textbf{Empirical scaling of unit-cell volume with system size.}
Two-dimensional histogram of unit-cell volume and full-cell heavy-atom count for 454,555 unique structures in the source crystal corpus.
Shading denotes the number of structures in each logarithmically spaced bin.
The blue line marks the size-conditioned lattice-prior centre, \(V_\star=17.784423N\).
Both axes are logarithmic.
}

\newcommand{\captionExtendedDataThree}{%
\textbf{Three-model performance on the shared 799-target benchmark.}
\textbf{a}, Target-best conformer RMSD, relative volume error, Hencky shape error and packing RMSD, together with pooled candidate-level severe-overlap prevalence.
Continuous metrics are minimized independently within each candidate pool and averaged over targets.
\textbf{b}, Hit rates by crystal system (left) and chemical class (right); a hit requires at least one clash-free candidate with packing RMSD below \(2\,\text{\AA}\).
SALA and Packora-L generate 50 candidates per target; CLARI uses target-specific pools of 50--3,150 candidates.
}

\newcommand{\captionExtendedDataFour}{%
\textbf{Ablation effects across molecular and crystallographic complexity.}
The full model and seven separately trained variants are stratified by ASU heavy-atom count (\textbf{a}), full-cell heavy-atom count (\textbf{b}), symmetry-operation count (\textbf{c}), ASU component count (\textbf{d}), molecular-component copies in the cell (\textbf{e}) and maximum rotatable-bond count among ASU components (\textbf{f}).
Columns show conformer RMSD, relative volume error, Hencky shape error, packing RMSD and samples with severe overlaps.
Each configuration uses the same candidate budget per target.
Continuous metrics are mean target-best values over targets evaluable across all configurations; overlap prevalence pools candidate-level outcomes.
Points denote means or percentages, and vertical dashed lines mark full-model values.
}

%% file: sections/abstract.tex
Molecular crystal packing shapes properties from drug bioavailability to charge transport, yet generating realistic structures requires coordinating molecular flexibility, intermolecular interactions and symmetry.
Here we introduce SALA, a flow-matching model built on state--context separation: it evolves only asymmetric-unit (ASU) atomic coordinates and symmetry-compatible lattice parameters while predicting their updates from the full periodic environment.
Conditioned on molecular graphs and a space group, SALA generates molecular conformation, lattice geometry and packing end to end while preserving the specified symmetry.
Across 799 held-out crystals spanning four chemical classes and all seven crystal systems, SALA achieves a 72.0\% hit rate from 50 candidates per target, compared with 8.1\% for an atomistic full-cell generative baseline.
Mean best-candidate packing root-mean-square deviation decreases from 4.48 to 1.70~\AA, with the advantage widening as system size and symmetry multiplicity increase.
SALA thus enables end-to-end generation of complex, symmetry-constrained molecular crystals, with potential applications in CSP, structure-based pharmaceutical polymorph design and organic molecular materials discovery.

%% file: sections/introduction.tex
\section*{Introduction}

The properties of molecular solids depend on both their constituent molecules and how those molecules pack, with consequences ranging from drug bioavailability to charge transport in organic semiconductors~\citep{cruzcabeza2015importance,coropceanu2007charge,johal2025evolutionary,lu2026metallic}.
The balance between intramolecular conformational preferences and intermolecular non-covalent interactions can allow a single compound to adopt distinct crystal structures, giving rise to polymorphism~\citep{nangia2008conformational,cruzcabeza2015importance,hall2026reality}.
Crystal structure prediction (CSP) seeks to identify these structures computationally, typically through candidate generation followed by structural relaxation and energy ranking~\citep{price2014predicting,zhou2025robust}.
Because accurate relaxation and ranking are often computational bottlenecks, the candidate pool determines both the packing arrangements accessible to the workflow and the cost of evaluating them~\citep{hoja2019reliable,nikhar2022firstprinciples,galanakis2024rapid}.
An effective generator must therefore cover plausible packing arrangements while limiting the number of candidates passed to downstream calculations.

Generating a molecular crystal requires searching a coupled, high-dimensional space of molecular conformations, lattice geometries and intermolecular arrangements~\citep{habgood2015flexibility}.
Conformational changes alter the contacts a molecule can form, while those contacts help determine its conformation, packing and periodic lattice.
The conformations of independent molecules, their positions and orientations, and the unit-cell parameters must consequently be determined together.
Crystallographic symmetry substantially reduces this search space. 
A space group specifies the symmetry operations---rotations, reflections, inversions and translations, including their combinations---under which a periodic structure is invariant.
Choosing a space group and its crystallographic setting constrains the cell angles and relationships between cell lengths, and organizes atoms into symmetry-equivalent sets, or orbits.
The asymmetric unit (ASU) contains the independent representatives from which the full cell can be reconstructed by applying these symmetry operations~\citep{case2016quasirandom,jiao2024diffcsppp}.
Space-group constraints thus provide both structural validity conditions and a reduction in dimensionality: a cell containing hundreds of atoms can be described by tens of independent atoms and a small number of lattice parameters (Fig.~\ref{fig:overview}a--c).

Conventional CSP methods exploit this reduction by selecting a space group and sampling molecular conformations, positions, orientations and compatible cell parameters~\citep{habgood2015flexibility,case2016quasirandom}.
Random or systematic sampling can then produce candidates that obey the selected symmetry before relaxation and ranking.
Nevertheless, finding relevant packing arrangements becomes increasingly difficult as molecular flexibility and compositional complexity increase, as highlighted by the seventh CSP blind test~\citep{hunnisett2024generation}.

Generative models offer a complementary route by learning structural distributions from crystal data~\citep{cheng2026aimaterials,wu2026siamese}.
For inorganic crystals, diffusion and flow-matching models generate atomic coordinates and lattices, and can also generate atomic species for de novo materials design~\citep{diffcsp,flowmm,luo2025crystalflow,zeni2025mattergen,okabe2026scigen}.
Text-guided diffusion and language models provide additional routes to crystal generation~\citep{park2025chemeleon,ahlawat2026llamat}.
Other developments incorporate valence-balance constraints or reinforcement-learning guidance to improve chemical validity, stability and diversity~\citep{cheng2026crysvcd,park2026chemeleon2}.
Crystallographic symmetry is incorporated by both established evolutionary searches and newer generative models.
Symmetry-guided evolutionary generators improve initialization and search efficiency in inorganic CSP~\citep{han2025symmetry}; DiffCSP++ constrains lattice geometry and Wyckoff coordinates, SymmCD models crystallographically independent atoms and their site symmetries, and WyckoffDiff generates discrete descriptions based on Wyckoff positions~\citep{jiao2024diffcsppp,levy2025symmcd,ekstrom2025wyckoffdiff}.

Generative modelling of organic molecular crystals is a newer and still emerging field.
Although early methods impose strong assumptions on molecular geometry or periodicity, recent work has focused primarily on all-atom generation.
Rigid-body methods such as MolCrystalFlow place supplied conformers without changing their internal geometry~\citep{molcrystalflow}.
OXtal allows conformational changes within molecular clusters but does not explicitly generate a periodic unit cell~\citep{oxtal}.
All-atom methods including PackFlow, CLARI, Packora and OrgFlow instead jointly generate molecular geometry, atomic positions and the lattice within an explicit periodic unit cell~\citep{packflow2026,clari,packora2026,orgflow2026}.

Transferring symmetry-aware generation to molecular crystals introduces an additional chemical requirement: symmetry-related atomic orbits must assemble into coherent molecules with plausible bond lengths, angles and torsions, while their preferred conformations depend on contacts with neighbouring molecules, including symmetry-related copies~\citep{jiao2024diffcsppp,levy2025symmcd,nangia2008conformational,habgood2015flexibility}.
A full-cell representation exposes these periodic interactions but, when symmetry-related atoms are generated independently, leaves exact orbit relationships to be learned.
Conversely, an ASU-only representation is compact but, without reconstructing its periodic surroundings, withholds the contact environment that shapes conformation and lattice geometry.
This tension grows with symmetry multiplicity, molecular flexibility and compositional complexity, motivating a symmetry-constrained generative state whose updates are predicted from the surrounding crystal environment.

Here we introduce SALA (Symmetry-Aware Lattice and Asymmetric-unit generation), a flow-matching model built on state--context separation (Fig.~\ref{fig:overview}d)~\citep{lipman2023flow}.
Given molecular graphs, ASU composition and a space group, SALA generates heavy-atom ASU coordinates and symmetry-compatible lattice parameters end to end, jointly determining molecular conformation, lattice geometry and packing.
SALA evolves only these independent variables; at each model evaluation, it reconstructs the full cell, predicts updates from the complete periodic environment and maps symmetry-related predictions back to their shared ASU representatives.
Within its general-position representation, the specified symmetry is preserved throughout sampling.
By specifying the space group and ASU composition, SALA can target a prescribed symmetry and $Z'$ or generate across plausible settings before structural relaxation and energetic ranking.
Across 799 held-out crystals spanning four chemical classes and all seven crystal systems, SALA achieves a 72.0\% hit rate from 50 candidates per target, compared with 8.1\% for an atomistic full-cell generative baseline.
Mean target-best packing RMSD decreases from 4.48 to 1.70~\AA, with SALA's advantage widening as system size and symmetry multiplicity increase.
Trajectory analyses and ablations reveal hierarchical structural organization and support the complementary roles of symmetry-constrained states, full-cell context and structure-aware design.
SALA thus enables symmetry-consistent generation of complex molecular crystals by separating constrained generative variables from the full periodic context needed to organize them.

%% file: sections/results.tex
\section*{Results}

\subsection*{Coupling symmetry-constrained states to full-cell context}

\input{sections/figure_1}

Within the flow-matching framework, SALA evolves a compact ASU--lattice state while using the full cell as the context for predicting its updates (Fig.~\ref{fig:overview}d).
Conditioned on the ASU chemical graphs and composition and a specified space group, the model jointly generates ASU Cartesian coordinates and a compatible lattice (Extended Data Fig.~\ref{fig:extended-data-1}).
SALA therefore assigns coordinates only to ASU atoms and retains only the independent lattice parameters compatible with the specified setting (Fig.~\ref{fig:overview}a--c).
Symmetry determines all remaining atomic coordinates, propagating each ASU and lattice update throughout the reconstructed crystal.

The evolving state comprises ASU coordinates $X_t$ and constrained lattice coordinates $q_t$, which a setting-specific decoder maps to a positive-volume lattice while fixing inactive components.
Sampling begins from a joint ASU--lattice prior: normalized ASU coordinates follow a standard Gaussian, whereas active lattice variables follow a size-conditioned Gaussian whose central volume is calibrated from the relationship between unit-cell volume and full-cell heavy-atom count (Extended Data Fig.~\ref{fig:extended-data-2}).
Flow matching learns the joint velocity field that transports this ASU--lattice prior to the data distribution~\citep{lipman2023flow,liu2023rectified}.
ASU-coordinate updates jointly control molecular conformation, component placement and relative spatial relationships among molecules, whereas lattice updates control the spatial relationships and periodicity of their symmetry- and translation-related images.
Predicting their coupled evolution therefore requires both molecular geometry and the surrounding network of non-covalent contacts in the same periodic environment.

The Gaussian prior state, interpolating states and numerical integration remain confined to the independent ASU and lattice variables.
At each model evaluation, OrbitExpand applies the space-group operations to the ASU atoms to construct the full-cell atoms, which enter a transformer together with three lattice tokens~\citep{vaswani2017attention,peebles2023dit}.
OrbitPullback applies the inverse group action to the full-cell atom-head vectors and averages them over each orbit, returning image-wise proposals to a common ASU frame; lattice predictions are likewise restricted to active components.
OrbitExpand then reconstructs the complete heavy-atom unit cell from the updated ASU and lattice, preserving the specified symmetry without assigning independent flow states to its images.

Whereas the unit cell is the translational repeat unit, a $3\times3\times3$ periodic expansion exposes short-range interactions across its boundaries.
Pair biases encode covalent connectivity within molecular components and non-covalent relations between periodic neighbours separated by less than $8\,\text{\AA}$, providing heuristic representations of intra- and intermolecular interactions, respectively.
No supercell atoms are maintained as additional tokens; instead, each periodic-neighbour bias enters the attention between the corresponding unit-cell atom tokens.
During training, random selection among symmetry-equivalent ASU component representatives serves as data augmentation, reducing sensitivity to representative choice; self-conditioning additionally supplies the preceding clean-endpoint prediction~\citep{chen2023analog}.
SALA thus combines a compact integrated state with full-cell attention and periodic interaction context.

\subsection*{SALA generates molecular crystals across chemical classes and crystal systems}

SALA can generate molecular crystal candidates for single-component, multicomponent, ionic and metal-containing systems, spanning multiple crystal systems and diverse molecular geometries and packing arrangements (Fig.~\ref{fig:performance}a).
Successful generation requires SALA to coordinate molecular conformation, lattice geometry and intermolecular packing.
We evaluated these three coupled structural levels using five metrics (Fig.~\ref{fig:performance}b,c; see Methods, ``Structural evaluation metrics''): conformer root-mean-square deviation (RMSD) for molecular conformation; relative volume and Hencky shape errors for lattice size and shape; and packing RMSD and the prevalence of severe atomic overlaps for intermolecular packing.
Packing RMSD compares assembled $2\times2\times2$ periodic supercells, thereby accounting for molecular packing across unit-cell boundaries.
For each continuous metric, we report target-best errors across candidate pools; a reference is counted as a hit when its pool contains at least one clash-free candidate with a packing RMSD below 2~\AA.

Table~\ref{tab:baselines} compares the representations and generative capabilities of recent molecular-crystal models.
For quantitative benchmarking, we selected CLARI~\citep{clari} and Packora~\citep{packora2026}, the least constrained end-to-end baselines that, like SALA, jointly generate molecular conformation, lattice and packing.
The held-out evaluation set comprises 799 crystals spanning four chemical categories, all seven crystal systems and broad variation in molecular and crystallographic complexity.
The main text focuses on CLARI; Packora-L is included in Extended Data Fig.~\ref{fig:extended-data-3}.

\input{sections/baseline_table}

Across the 799 held-out targets, SALA achieved 575 hits (72.0\%), compared with 65 (8.1\%) for CLARI.
SALA's higher hit rate coincided with a marked improvement in periodic packing: mean target-best packing RMSD decreased from 4.48 to 1.70~\AA, while candidate-level severe-overlap prevalence fell from 92.2\% to 34.8\% (Fig.~\ref{fig:performance}c).
Mean target-best conformer RMSD decreased from 3.33 to 0.52~\AA, and mean target-best Hencky shape error decreased from 0.072 to 0.010, showing that SALA generated coherent molecular conformations and cell shapes together with improved intermolecular arrangements.
CLARI achieved a lower mean target-best relative volume error of 0.14\%, compared with 0.40\% for SALA; however, its higher mean target-best Hencky shape and packing errors show that matching cell volume alone is insufficient to produce a hit.

\input{sections/figure_2}

SALA's hit-rate advantage persisted across chemical and crystallographic categories.
Hit rates ranged from 68.3\% for multicomponent crystals to 83.6\% for ionic crystals, with rates of 73.6\% and 68.5\% for single-component and metal-containing systems, respectively (Fig.~\ref{fig:performance}e).
SALA also achieved higher hit rates than CLARI in all seven represented crystal systems, reaching 93.9\% for triclinic and 43.6\% for trigonal targets and producing one hit among the 11 cubic targets (Fig.~\ref{fig:performance}d).
These broad hit-rate gains prompted us to examine how SALA's advantage scales with the size of the independent molecular assembly and the symmetry multiplicity of the full cell.

\subsection*{SALA's packing advantage widens with molecular and crystallographic complexity}

To determine how SALA's packing advantage scales with system complexity, we considered four complementary axes spanning the independent molecular assembly and the symmetry-expanded crystal: ASU heavy-atom count, full-cell heavy-atom count, molecular-component copies in the cell and symmetry-operation count (Fig.~\ref{fig:complexity}a).
These axes capture, respectively, the number of independently generated atomic coordinates, the overall scale of the packing environment, molecular multiplicity within the cell and crystallographic symmetry multiplicity.

SALA's packing error increased more slowly than CLARI's across all three measures of system scale: ASU size, full-cell size and molecular-component copies.
From 8--16 to 70--100 ASU heavy atoms, mean target-best packing RMSD rose from 0.83 to 2.48~\AA\ for SALA, compared with 2.60 to 6.32~\AA\ for CLARI.
Across full-cell sizes from 16--67 to at least 501 heavy atoms, the corresponding errors increased from 0.78 to 2.86~\AA\ and from 2.31 to 9.23~\AA, respectively.
Likewise, from 1--4 to at least 25 molecular-component copies per cell, packing RMSD increased from 0.85 to 3.62~\AA\ for SALA and from 2.47 to 10.32~\AA\ for CLARI.
Together, these trends show that SALA's advantage grows as both the independently generated assembly and the periodic packing environment become more complex.

The separation was strongest along the crystallographic-symmetry axis.
From 3--4 to 12--18 symmetry operations, SALA's mean packing RMSD increased only from 1.67 to 2.06~\AA, whereas CLARI's increased from 3.01 to 7.42~\AA.
In the highest-multiplicity bin, with at least 24 operations, the gap reached 6.26~\AA\ (6.00 versus 12.26~\AA).
This pattern is consistent with SALA's symmetry-constrained representation, which derives every symmetry-related image by deterministic orbit expansion from common ASU representatives rather than assigning each image an independent generative state.

\input{sections/figure_3}

Joint stratification showed that this symmetry-dependent advantage was not explained solely by larger molecules or cells (Fig.~\ref{fig:complexity}b).
Within comparable ASU-size, full-cell-size and component-copy groups, the mean CLARI-minus-SALA packing-error difference generally increased with symmetry multiplicity.
For ASUs containing 60--100 heavy atoms, for example, the gap widened from 1.37~\AA\ at 3--4 operations to 4.88~\AA\ at 6--9 operations and 8.32~\AA\ at 12 or more operations, with 41 targets in the highest-symmetry group.
The strongest advantages therefore emerged when large independent molecular assemblies were reproduced across many symmetry-related images.
These results link SALA's scaling advantage to its separation of a symmetry-constrained state from full-cell context.
To probe the generative process underlying this advantage, we examined SALA's evolving structural predictions and internal attention patterns.

\subsection*{Trajectory-resolved attention reveals how SALA organizes crystal structure}

We first resolved the temporal order in which SALA organized cell geometry, molecular conformation and intermolecular packing.
Using each trajectory's final output as an internal reference, we quantified the normalized progress of its clean-endpoint predictions (Fig.~\ref{fig:dynamics}a).
Halfway through generation, mean progress had reached 0.83 for the cell, compared with 0.48 for conformation and 0.53 for packing, while intermolecular contact-set overlap remained 0.23.
SALA therefore established the cell geometry early, then refined molecular conformations and relative placements within this emerging periodic framework before stabilizing the contact network.

\input{sections/figure_4}

We next asked whether SALA's internal computation tracked this structural sequence.
As generation progressed, attention became increasingly enriched for covalent pairs and for intermolecular pairs that formed close contacts in the final structure, while enrichment declined for distant pairs and atom-to-cell relations (Fig.~\ref{fig:dynamics}b).
The pair bias alone reproduced part of this redistribution: enrichment for final intermolecular contacts increased by 13.5\% in the bias-only distribution, compared with 20.6\% in full attention (Fig.~\ref{fig:dynamics}c).
The larger full-attention increase, particularly for covalent pairs, indicates that evolving content representations refine the static relation encoding rather than merely reproducing it.

Distance--time maps localized this redistribution in geometric space, revealing a pronounced shift in attention towards shorter separations for non-bonded intramolecular pairs (Fig.~\ref{fig:dynamics}d).
At the level of individual intermolecular pairs, attention increased as the atoms approached and established persistent contacts (Fig.~\ref{fig:dynamics}e).
Together, the trajectory and attention analyses show that SALA's computational focus tracks a hierarchical organization process: the cell forms first, molecular geometry and packing are then refined, and attention increasingly concentrates on the covalent structure and emerging contact network.

\subsection*{Ablations support SALA's state--context framework and explain its design principles}

To identify the sources of SALA's performance, we evaluated seven separately trained ablations spanning the state--context framework and its structure-aware design choices (Fig.~\ref{fig:ablation}).
The first group tested the framework that couples a symmetry-constrained state to full-cell context.
Restricting the transformer context to the ASU increased mean target-best packing RMSD from 1.64 to 5.12~\AA\ and the clash rate from 34.9\% to 77.2\%, showing that a compact independent state alone does not provide sufficient information to organize periodic packing.
Removing the orbit operations increased conformer RMSD from 0.50 to 4.18~\AA, packing RMSD to 3.73~\AA\ and the clash rate to 99.3\%.
Allowing all six lattice coordinates to vary without space-group projection likewise increased Hencky shape error from 0.009 to 0.063 and packing RMSD to 2.95~\AA.
These complementary degradations support the central framework: accurate generation requires full-cell interaction context, a symmetry-constrained state and an orbit-consistent map between them.

\input{sections/figure_5}

The remaining ablations clarify the roles of SALA's structure-aware heuristics.
Replacing the size-conditioned lattice prior with a standard-normal prior increased Hencky shape error to 0.220 and packing RMSD to 5.23~\AA, consistent with the prior supplying a system-size-appropriate starting scale for coupled lattice and packing generation.
Removing pair bias increased conformer RMSD to 4.82~\AA, packing RMSD to 3.07~\AA\ and the clash rate to 92.8\%.
This multi-channel degradation complements the trajectory analysis: explicit covalent and periodic-neighbour relations help organize both intramolecular geometry and intermolecular contacts.
Removing self-conditioning increased conformer and packing RMSD to 1.86 and 3.03~\AA, respectively, consistent with preceding endpoint estimates supporting iterative refinement.
Finally, disabling symmetry-equivalent representative augmentation increased conformer RMSD to 1.55~\AA\ and packing RMSD to 3.05~\AA, supporting its intended role in reducing sensitivity to the chosen ASU representation.

Complexity-stratified results connected these contributions to the scaling trends in Fig.~\ref{fig:complexity} (Extended Data Fig.~\ref{fig:extended-data-4}).
The packing penalty from restricting context to the ASU widened with full-cell size, while removing orbit operations or pair bias produced large conformational errors across the size range.
Thus, full-cell context and its symmetry- and relation-aware coupling to the independent state become increasingly consequential as more atoms and molecular images must be coordinated.

No single component accounted for all structural channels: perturbing context, symmetry handling, lattice construction or learned relations propagated across conformation, cell geometry and packing.
The full model achieved the most balanced accuracy across these outputs, supporting SALA's central principle that crystallographic constraints define a compact independent state, while the expanded periodic environment provides the context required to organize it into a coherent molecular crystal.

%% file: sections/figure_1.tex
\par\bigskip
\noindent\begin{minipage}{\linewidth}
\makebox[\linewidth][c]{\includegraphics{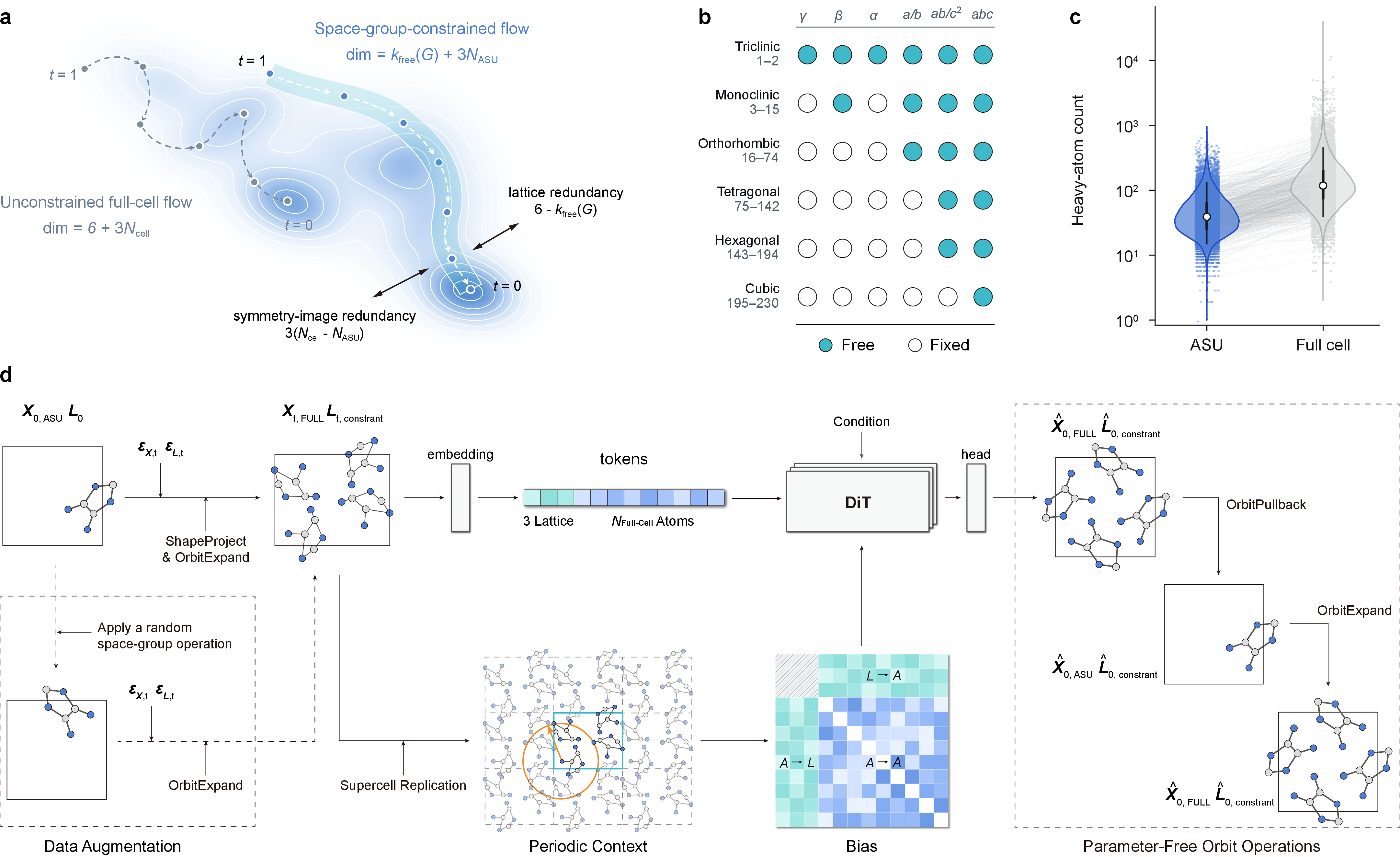}}
\captionof{figure}{\captionFigOne}\label{fig:overview}
\par\vspace{0.6em}\hrule height0.5pt
\end{minipage}
\par\bigskip

%% file: sections/baseline_table.tex
\noindent\begin{minipage}{\linewidth}
\captionof{table}{\textbf{Representations and generative capabilities of molecular-crystal models.}}
\label{tab:baselines}
\vspace{0.4em}
\begingroup
\footnotesize
\setlength{\tabcolsep}{2.5pt}
\renewcommand{\arraystretch}{1.25}
\begin{tabular*}{\linewidth}{@{\extracolsep{\fill}}p{0.17\linewidth}p{0.17\linewidth}p{0.16\linewidth}cccc@{}}
\toprule
Method & Input & Coordinate state & \multicolumn{4}{c}{Generative capability} \\
\cmidrule(lr){4-7}
 & & & \shortstack[c]{Conformation\\\strut} & \shortstack[c]{Lattice\\\strut} & \shortstack[c]{Packing\\\strut} & \shortstack[c]{Multiple chemical\\classes} \\
\midrule
OXtal~\citep{oxtal} & Graphs + molecule counts in cell & Molecular-crop atoms & \ding{51} & \ding{55} & \ding{51} & \ding{55} \\
MolCrystalFlow~\citep{molcrystalflow} & Graphs + molecule counts in cell + 3D conformers & Rigid-molecule poses & \ding{55} & \ding{51} & \ding{51} & \ding{55} \\
PackFlow~\citep{packflow2026} & Graphs + molecule counts in cell & Full-cell atoms & \ding{51} & \ding{51} & \ding{51} & \ding{55} \\
CLARI~\citep{clari} & Graphs + molecule counts in cell & Full-cell atoms & \ding{51} & \ding{51} & \ding{51} & \ding{51} \\
Packora~\citep{packora2026} & Graphs + molecule counts in cell & Full-cell atoms & \ding{51} & \ding{51} & \ding{51} & \ding{51} \\
\midrule
\textbf{SALA} & ASU graphs + space group & \textbf{ASU atoms} & \ding{51} & \ding{51} & \ding{51} & \ding{51} \\
\bottomrule
\end{tabular*}\par
\endgroup
\vspace{0.4em}
\par\noindent\footnotesize
\ding{51}, supported; \ding{55}, not natively generated.
Coordinate state identifies the independently generated atomic coordinates or molecular poses; SALA reconstructs symmetry-related cell atoms from the ASU.
Conformation denotes variable intramolecular geometry; lattice denotes explicit generation of cell parameters.
Multiple chemical classes denotes native support beyond a single neutral molecular-crystal class.
\end{minipage}
\medskip

%% file: sections/figure_2.tex
\par\bigskip
\noindent\begin{minipage}{\linewidth}
\makebox[\linewidth][c]{\includegraphics{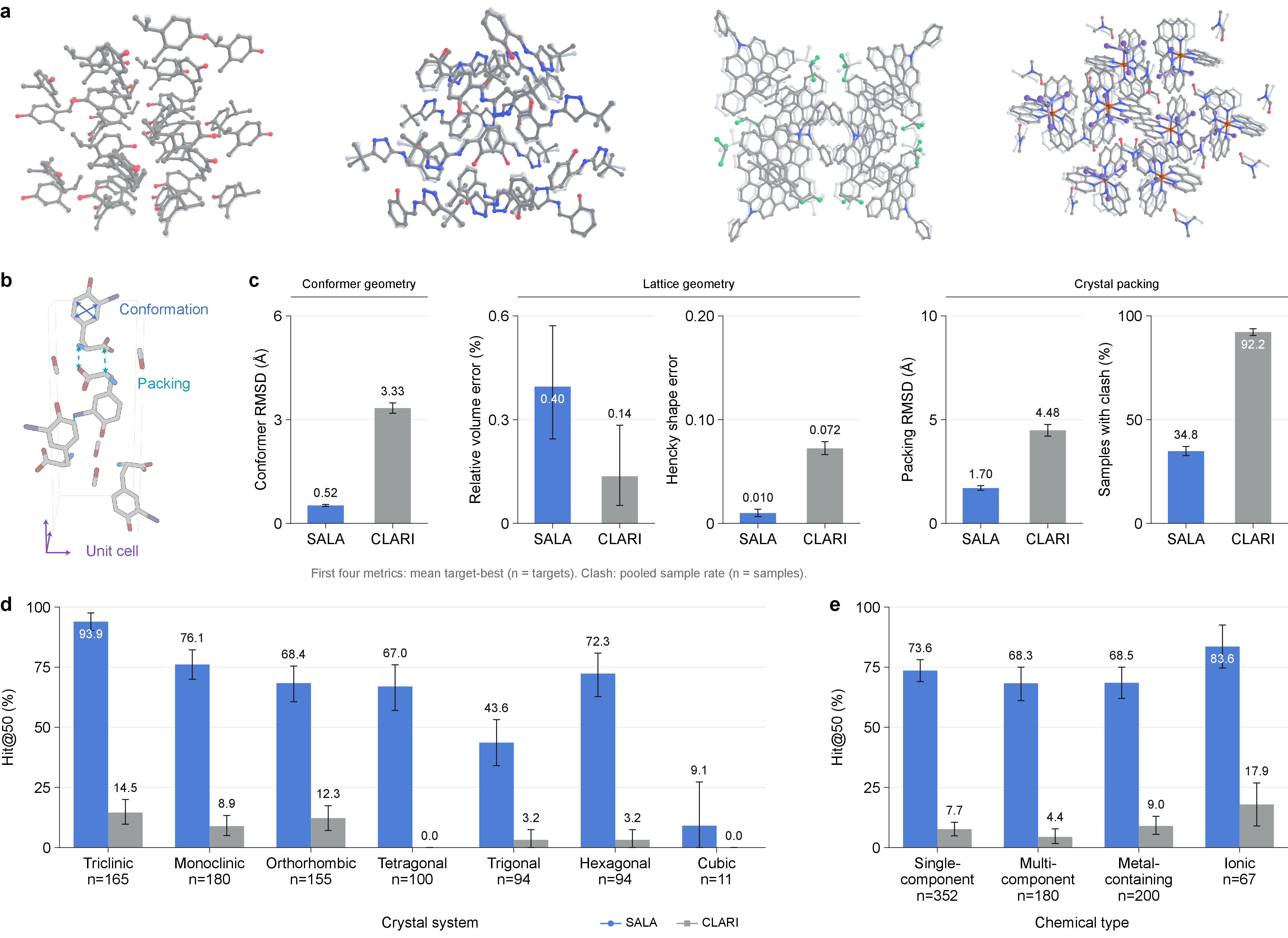}}
\captionof{figure}{\captionFigTwo}\label{fig:performance}
\par\vspace{0.6em}\hrule height0.5pt
\end{minipage}
\par\bigskip

%% file: sections/figure_3.tex
\par\bigskip
\noindent\begin{minipage}{\linewidth}
\makebox[\linewidth][c]{\includegraphics{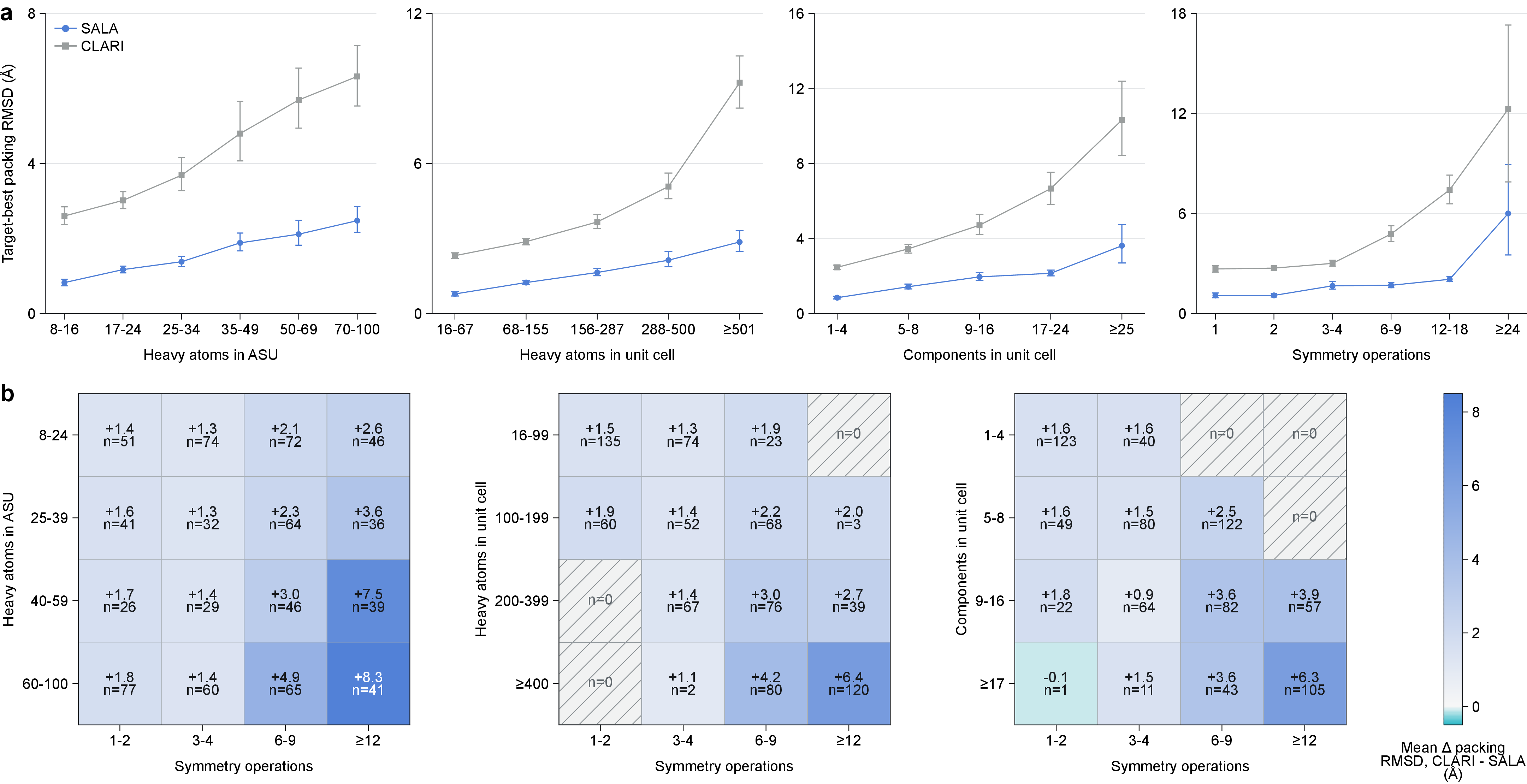}}
\captionof{figure}{\captionFigThree}\label{fig:complexity}
\par\vspace{0.6em}\hrule height0.5pt
\end{minipage}
\par\bigskip

%% file: sections/figure_4.tex
\par\bigskip
\noindent\begin{minipage}{\linewidth}
\makebox[\linewidth][c]{\includegraphics{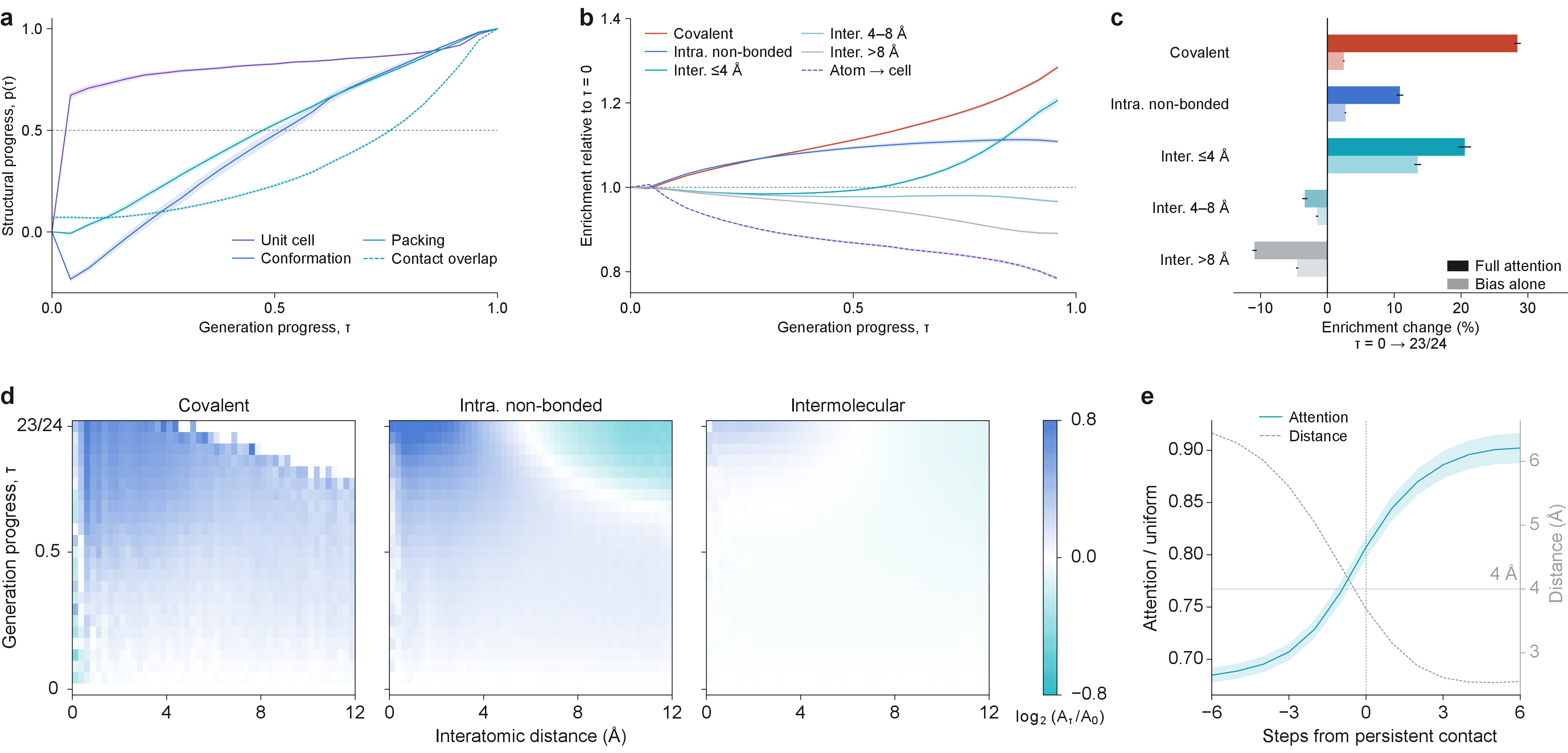}}
\captionof{figure}{\captionFigFour}\label{fig:dynamics}
\par\vspace{0.6em}\hrule height0.5pt
\end{minipage}
\par\bigskip

%% file: sections/figure_5.tex
\par\bigskip
\noindent\begin{minipage}{\linewidth}
\makebox[\linewidth][c]{\includegraphics{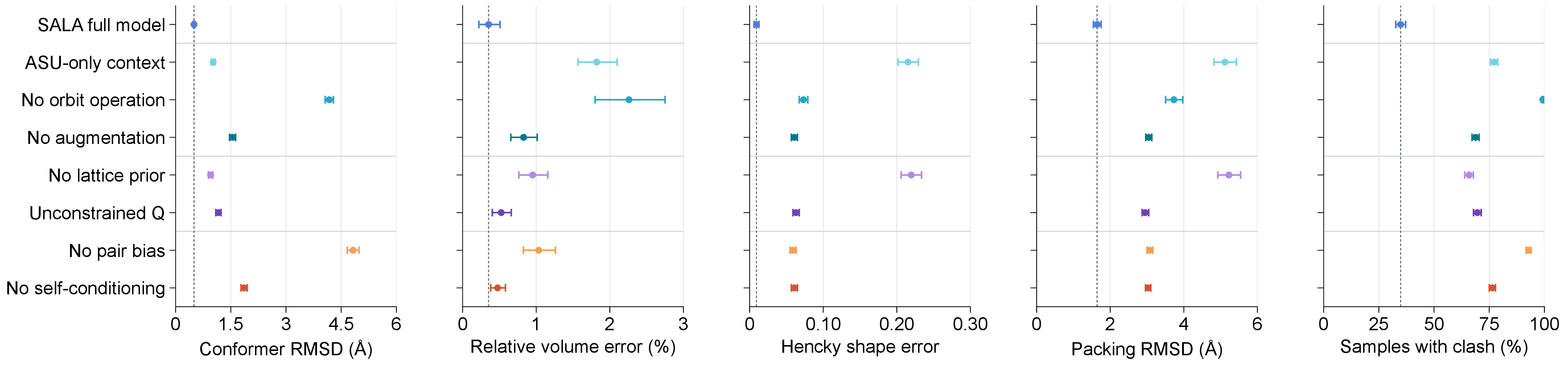}}
\captionof{figure}{\captionFigFive}\label{fig:ablation}
\par\vspace{0.6em}\hrule height0.5pt
\end{minipage}
\par\bigskip

%% file: sections/discussion.tex
\section*{Discussion}

SALA addresses molecular crystal generation by separating the coordinates that must evolve independently from the context needed to model their interactions.
The ASU and admissible lattice parameters form a compact generative state; deterministic orbit expansion reconstructs the full cell for contextual computation; and orbit pullback returns the predicted updates to the independent variables.
This organization preserves the encoded crystallographic relationships while allowing molecular conformation, cell geometry and packing to evolve jointly.
Its empirical signature is strongest where the representation should matter most: the advantage over full-cell generation widens as symmetry multiplicity increases.
Trajectory analysis shows how the lattice, molecular geometry and contact network organize in sequence, while ablations identify expanded-cell context and constrained lattice evolution as major sources of packing accuracy.
Together, these results show that separating the constrained state from full-cell context enables SALA to preserve crystallographic symmetry while modelling the interactions that govern molecular-crystal packing.

This distinction enables controllable candidate generation.
Specifying the space group and ASU composition, and hence $Z'$, focuses sampling on a physically meaningful structural family while retaining molecular flexibility.
SALA can therefore generate within a known space group, search across plausible groups or target a particular $Z'$ before relaxation and energetic ranking~\citep{price2014predicting}.
This control also provides a natural interface to experimental constraints, including spectroscopic signatures~\citep{paruzzo2018chemical}, and to energy or free-energy models that select stable candidates~\citep{hoja2019reliable,firaha2023realworld,pulido2017functional}.
The broader opportunity is a generate--evaluate workflow in which crystallographic and experimental knowledge shapes candidate proposal, while accurate physical models rank the resulting polymorph landscape.

The current representation is limited to discrete molecular components whose heavy atoms occupy general Wyckoff positions.
It cannot natively represent molecules on special positions, which require site stabilizers and symmetry-induced atom permutations; such structures can currently be accommodated only by lowering the symmetry to $P1$, forfeiting the corresponding crystallographic constraints.
The component-based molecular graph also excludes periodically connected topologies, including metal--organic frameworks, covalent organic frameworks and polymers.
Because the supplied chemical graph fixes atomic connectivity and chemical identity, the model does not explore changes in stereochemistry or proton-transfer states.
Extending the state representation to these degrees of freedom would broaden SALA's chemical and crystallographic scope, while sparse or local full-cell computation could improve scaling to larger cells.

By showing that symmetry-constrained ASU and lattice flows can exploit full-cell interactions to generate diverse crystal structures, SALA establishes a foundation for symmetry-constrained generative modelling of molecular solids.
Coupling this framework to physical ranking and experimental feedback could enable iterative exploration of polymorph landscapes, directing generation beyond structural validity towards thermodynamic accessibility, synthesizability and target solid-state properties.
In pharmaceutical development, such workflows could identify polymorphs with favourable stability, manufacturability and bioavailability; in organic materials discovery, they could support the inverse design of crystal packings that control charge transport, optical response and other collective functions~\citep{smalley2026structural}.
Extending SALA with property conditioning, physical ranking and experimental feedback could ultimately support the joint optimization of molecular crystal structure and function.

%% file: sections/methods.tex
\section*{Methods}
\label{sec:methods}

\subsection*{Crystallographic state and symmetry operations}

SALA generates a periodic molecular crystal conditioned on an ASU chemical graph, its composition and a specified space group. The independent state contains heavy-atom ASU Cartesian coordinates $X$ and a six-slot lattice coordinate $q$; experimental coordinates and lattices are used only as training endpoints and evaluation references. The current representation assigns one ASU atom to each general-position orbit and excludes reduced-multiplicity sites and intersecting representative orbits. CIF parsing, molecular reconstruction, chemical features and filtering are described in the Supplementary Information~\cite{hall1991cif,groom2016csd}.

We use row-vector coordinates $x=fA$, with lattice metric $G=AA^{\mathsf T}$. A fractional symmetry operation $(R_g,t_g)$ acts as $f\mapsto fR_g^{\mathsf T}+t_g$, and a compatible lattice satisfies $R_g^{\mathsf T}GR_g=G$. The Hall setting selects an analytic lattice template and active mask $M_G$~\cite{togo2024spglib}. Active length coordinates are logarithmic; equal lengths and fixed angular relations are imposed by the decoder, which maps $q$ to a canonical lower-triangular, positive-volume lattice. This yields 6, 4, 3, 2, 2, 2 and 1 independent coordinates for triclinic, monoclinic, orthorhombic, tetragonal, trigonal, hexagonal and cubic cells, respectively. Setting-specific templates and decoding equations are given in the Supplementary Information.

OrbitExpand deterministically constructs the full-cell coordinates from the independent ASU state,
\begin{equation}
\mathcal E_A(X)_g=(XA^{-1}R_g^{\mathsf T}+t_g)A,
\label{eq:orbit-expansion}
\end{equation}
so general-position structures contain $N_{\mathrm{cell}}=|\mathcal G|N_{\mathrm{ASU}}$ contextual atoms but only $N_{\mathrm{ASU}}$ independent atomic states. After full-cell computation, OrbitPullback expresses each atom-head vector $u_{i,g}$ in the common ASU frame and averages over its orbit,
\begin{equation}
v_i^X=\frac{1}{|\mathcal G|}\sum_g u_{i,g}A^{-1}R_g^{-\mathsf T}A.
\label{eq:orbit-pullback}
\end{equation}
Lattice outputs are masked by $M_G$, and the updated independent state reconstructs the full periodic crystal. Intermolecular geometry uses an exact closest-image search with lattice-dependent anisotropic bounds; it does not instantiate additional supercell tokens. Implementation details and consistency checks are provided in the Supplementary Information.

\subsection*{Model architecture and flow-matching training}

The network processes three lattice tokens and one token per full-cell heavy atom. Two covalent message-passing layers encode the ASU graph before its features are broadcast to symmetry images; component features summarize composition, charge and size. A 24-block Crystal-DiT with width 1,024, 16 attention heads and SwiGLU width 2,816 updates the joint token set~\citep{shazeer2020glu,peebles2023dit}. Global conditioning includes flow time, space group, lattice template and atom and component counts.

Dense attention is augmented by pair biases encoding token type, covalent bond type, shared orbit, component relation and geometry. Intramolecular geometry uses unwrapped displacements. For different molecular instances, radial and fractional features use the closest periodic displacement, with the geometric contribution smoothly truncated at 8~\AA; relation biases and content attention remain global. The atom head is reduced to ASU velocities by OrbitPullback, while the three lattice tokens predict the masked lattice velocity.

The normalized data state is $y_0=(X_0/s_X,z_0)$, where $s_X=8\,\text{\AA}$ and $z_0=M_G\odot(q_0-\mu_G)/\sigma_G$. The lattice prior is centred at volume $17.784423N_{\mathrm{cell}}\,\text{\AA}^3$, with standard deviations 0.30 for active log-length coordinates and 0.20 for active angular coordinates. Rectified flow matching~\cite{lipman2023flow,liu2023rectified} uses
\begin{equation}
y_t=(1-t)y_0+t\epsilon,\qquad v^*=\epsilon-y_0,
\end{equation}
with standard-normal ASU noise and masked lattice noise. Times follow a clipped logit-normal distribution, and the network predicts the joint velocity and clean endpoint $\widehat y_0=y_t-tv_\theta$.

The formal objective combines orbit-expanded Cartesian velocity matching, active-coordinate lattice velocity matching and a late-trajectory contact--clash term:
\begin{equation}
\mathcal L=\mathcal L_{v,X}+\mathcal L_{v,q}+0.10\mathcal L_{\mathrm{contact+clash}}.
\end{equation}
The contact--clash term is active for $t\leq0.4$ and combines penetration penalties with Huber matching of reference intermolecular contacts within 5~\AA. Each crystal is normalized independently before batch averaging.

During training, symmetry-equivalent ASU representatives are sampled with probability 0.5, and self-conditioning supplies a detached preliminary endpoint estimate to half of the crystals~\cite{chen2023analog}. 
The model is trained on a single node with eight NVIDIA H20 GPUs (96 GB each) for 250,000 optimizer updates, using BF16 transformer computation, hybrid Muon--AdamW optimization and cosine decay after 3,000 warm-up steps~\citep{jordan2024muon,loshchilov2019adamw}. 
Sampling integrates from $t=1$ to 0 over 24 uniform intervals using an Euler initialization followed by second-order Adams--Bashforth updates, with one network evaluation per interval. 
Full architectural, objective and optimization details are provided in the Supplementary Information.

\subsection*{Structural evaluation metrics}

Five heavy-atom metrics quantify three coupled structural levels. For a molecular component $c$ containing $N_c\geq2$ heavy atoms with a chemistry-preserving correspondence, conformer RMSD is
\begin{equation}
r_c=\min_{Q\in\mathrm{SO}(3),\,b\in\mathbb R^3}
\left[\frac{1}{N_c}\sum_{i\in c}\left\|\widehat x_iQ+b-x_{\mathrm r,i}\right\|_2^2\right]^{1/2}.
\end{equation}
Each component is aligned independently by a proper Kabsch rotation~\citep{kabsch1976}. Let $\mathcal T$ denote the applicable chemical types and $\mathcal C_k$ the components of type $k$; the reported chemical-type-macro conformer RMSD is
\begin{equation}
E_{\mathrm{conf}}=\frac{1}{|\mathcal T|}\sum_{k\in\mathcal T}\frac{1}{|\mathcal C_k|}\sum_{c\in\mathcal C_k}r_c.
\end{equation}
Thus, repeated copies do not give a chemical type greater weight, and single-heavy-atom components are excluded.

Lattice scale is measured throughout by relative volume error,
\begin{equation}
E_{V,\mathrm{rel}}=\frac{|\widehat V-V_{\mathrm r}|}{V_{\mathrm r}}\times100\%.
\end{equation}
Lattice shape is separated from scale using $\overline G=G/V^{2/3}$. If $e_i=\tfrac12\log\lambda_i$ are the principal strains of $\overline G_{\mathrm r}^{-1/2}\overline G\,\overline G_{\mathrm r}^{-1/2}$, the Hencky shape error is
\begin{equation}
E_{\mathrm{shape}}=\left[\tfrac13\sum_{i=1}^3(e_i-\overline e)^2\right]^{1/2}.
\end{equation}
Comparisons retain the conditioned Hall-setting correspondence.

Intermolecular packing is assessed by packing RMSD and severe-overlap prevalence. After establishing a chemistry-preserving correspondence between the $N$ unit-cell atoms, the generated and reference $2\times2\times2$ supercells are constructed with their respective lattices as
\begin{equation}
\widehat P_{i,n}=\widehat x_i+n\widehat A,\qquad
P_{\mathrm r,i,n}=x_{\mathrm r,i}+nA_{\mathrm r},
\qquad n\in\{0,1\}^3.
\end{equation}
Flattening $(i,n)$ to $j=1,\ldots,8N$, packing RMSD is
\begin{equation}
E_{\mathrm{pack}}=\min_{Q\in\mathrm{SO}(3),\,b\in\mathbb R^3}
\left[\frac{1}{8N}\sum_{j=1}^{8N}\left\|\widehat P_jQ+b-P_{\mathrm r,j}\right\|_2^2\right]^{1/2}.
\end{equation}
Only one global proper alignment is fitted to the complete supercells, so lattice deformation, molecular placement and conformation remain in the error. A candidate has a severe overlap if the distance of any eligible intermolecular periodic pair is below $\min[d_{ij}^{\mathrm r},\max(0,r_i^{\mathrm{vdW}}+r_j^{\mathrm{vdW}}-2\,\text{\AA})]$. Periodic-search implementation details are provided in the Supplementary Information.

\subsection*{Experimental protocols}

The main benchmark uses 799 held-out crystals spanning all seven crystal systems: 352 single-component, 180 multicomponent, 67 ionic and 200 metal-containing targets. SALA and Packora-L each generate 50 candidates per target. Because CLARI-H is not conditioned on a specified space group, it receives additional sampling to cover crystallographic alternatives with the same order as the target space group: its target-specific budget is $50n_m$, where $n_m$ is the number of standard space groups with that order, yielding pools of 50--3,150 candidates. SALA is conditioned on the ASU graph and specified space group, and CLARI-H uses its official checkpoint. All five metrics are evaluated on raw generated heavy-atom structures.

Each continuous metric is minimized independently within each candidate pool and then averaged with equal weight over mutually evaluable targets. A target is a hit if any clash-free candidate has packing RMSD strictly below 2~\AA; severe-overlap prevalence is reported separately over candidates with known outcomes. Baseline preparation, candidate allocation and aggregation details are provided in the Supplementary Information.

Figure~\ref{fig:ablation} compares the full model with seven separately trained variants: ASU-only context, no orbit operations, no representative augmentation, no lattice prior, unconstrained lattice coordinates, no pair bias and no self-conditioning. Each configuration uses the same candidate budget per target. Continuous metrics use targets evaluable across all models; clash summaries pool known candidate outcomes from the full cohort. Extended Data Fig.~\ref{fig:extended-data-4} applies the same reductions within strata of ASU and cell size, symmetry multiplicity, component count and molecular flexibility. Ablation configurations, training schedules, source-data provenance and reproducibility records are provided in the Supplementary Information.

\subsection*{Generation dynamics and attention analyses}

For Fig.~\ref{fig:dynamics}, each target was sampled with ten deterministic seeds, and the state, velocity, clean-endpoint prediction and attention were recorded at each of 24 model evaluations. At generation progress $\tau=1-t$, every clean-endpoint prediction was compared with the final output of the same trajectory. The cell error is $\|L_\tau L_{\mathrm{final}}^{-1}-I\|_F/\sqrt{3}$. Conformation is measured by the root-mean-square change in all off-diagonal intramolecular distances, and packing by the root-mean-square change in minimum-image intermolecular distances for pairs within 6~\AA\ in the final structure, with distances capped at 12~\AA. For target $i$ and structural channel $m$, errors are averaged over seeds before normalization:
\begin{equation}
p_{i,m}(\tau)=1-\frac{\overline{\varepsilon}_{i,m}(\tau)}
{\overline{\varepsilon}_{i,m}(0)}.
\end{equation}
Intermolecular contact overlap is the Jaccard index
\begin{equation}
J_i(\tau)=
\frac{|C_i(\tau)\cap C_i(1)|}{|C_i(\tau)\cup C_i(1)|},
\end{equation}
where $C_i(\tau)$ contains intermolecular pairs separated by at most 4~\AA.

Attention probabilities are averaged over layers and heads. For a fixed pair class $C$ and $T$ valid tokens, enrichment is
\begin{equation}
E_C(\tau)=\frac{T}{|C|}\sum_{(i,j)\in C}a_{ij}(\tau),
\end{equation}
which measures mean pair attention relative to uniform attention over valid keys. Pair classes are defined by covalent connectivity, molecular-instance identity and final pair distance and remain fixed along each trajectory; Fig.~\ref{fig:dynamics}b normalizes each target's enrichment by its first-evaluation value. Bias-only attention retains the learned bias scaling and head mixing but removes query--key content logits before masking and softmax.

Distance--time maps instead group pairs by their instantaneous separation in 0.25~\AA\ bins from 0 to 12~\AA. The two directed attention entries are averaged, and colour reports the base-2 logarithm of each bin's mean attention relative to the corresponding bin at the first evaluation. A persistent-contact event is the first recorded evaluation after which a final-contact pair remains within 4~\AA\ for the rest of the trajectory; bidirectional attention and separation are aligned within six evaluations on either side. Curves and bars first average seeds within targets and then weight targets equally, whereas distance--time maps pool pair sums and counts. These diagnostics describe how computation is allocated as geometry evolves, but attention alone is not interpreted as causal evidence~\citep{jain2019attention}. Additional implementation details are provided in the Supplementary Information.

\subsection*{Statistical analysis}

Unless otherwise stated, error bars and shaded intervals in the main and Extended Data figures denote 95\% bootstrap intervals, defined by the 2.5th and 97.5th percentiles of 10,000 resamples. The target is the resampling unit for continuous metrics, hit rates, trajectory summaries and their stratified analyses. For severe-overlap prevalence and other summaries that pool candidate-level outcomes, whole target clusters are resampled so that all candidates associated with a sampled target remain together. Complexity analyses apply the same target-level reductions and resampling procedure within each stated stratum.

%% file: sections/declarations.tex
\section*{Data availability}

Crystal structures were obtained from the Cambridge Structural Database (CSD), which requires a licence for access. We accessed the CSD under an academic licence held by Peking University. Entry metadata were queried through the CSD Python API. From the processed collection, 799 crystals were selected as a held-out evaluation set to span chemical, crystallographic and system complexity. To prevent molecular-topology leakage, all remaining records sharing a molecular topology with any held-out target were excluded before model development; the resulting data were divided into training and validation sets in a 95:5 ratio. Owing to CSD licensing restrictions, the underlying CSD records cannot be redistributed by the authors and are available from the Cambridge Crystallographic Data Centre to licensed users. Data-processing and partitioning procedures are described in the Supplementary Information.

\section*{Code availability}

The source code and trained model checkpoints are available at \url{https://github.com/WENDI-CAI/SALA.git}.

\section*{Declaration of AI use}

During the preparation of this work, the authors used the Seed model (ByteDance) to assist with data processing, code development for reproducing baseline models used in the benchmarks, results visualization, and language editing and proofreading. The authors reviewed and edited all outputs and take full responsibility for the scientific content, experimental design, analysis and conclusions.

\section*{Acknowledgements}

This work was supported by the National Natural Science Foundation of China through the Young Scientists Fund-Type A (the former National Science Fund for Distinguished Young Scholars, NSFDYS) (Grant No. T262500040) and the General Program (Grant No. 22673006). We gratefully acknowledge Peking University Shenzhen Graduate School and the Shenzhen Government for start-up funding support, and the High-Performance Computing Platform of Peking University for providing computational resources. We thank the Seed STEM Fellows Program for providing computational resources and regular discussions. We thank Ningyi Lyu and Chengchun Liu for helpful discussions that improved our understanding and presentation of this work.

\section*{Competing interests}
The authors declare no competing interests.

%% file: sections/extended_data.tex
\begingroup
\renewcommand{\figurename}{Extended Data Fig.}
\setcounter{figure}{0}
\renewcommand{\theHfigure}{extended.\arabic{figure}}

\noindent\begin{minipage}{\linewidth}
\makebox[\linewidth][c]{\includegraphics{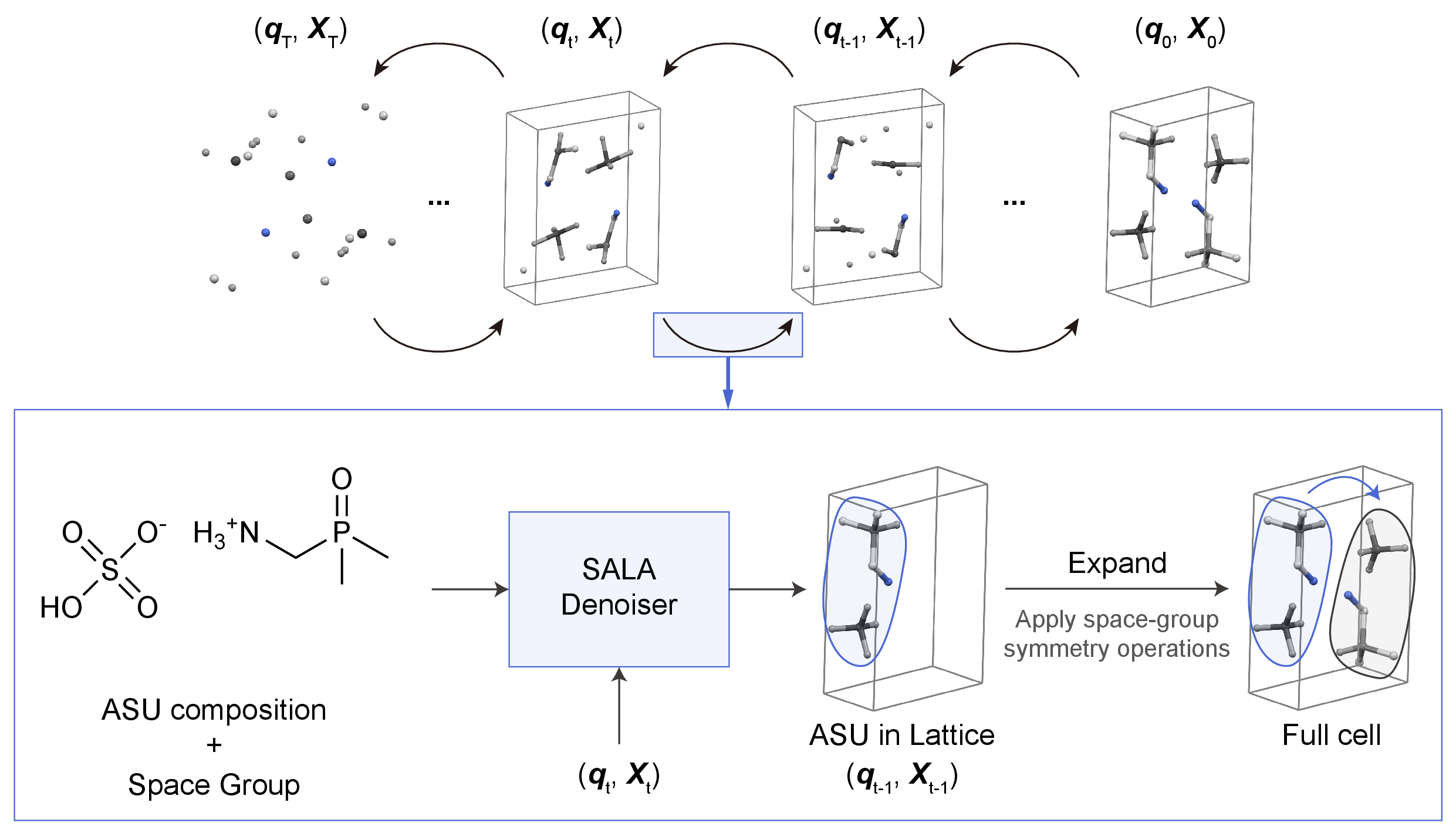}}
\captionof{figure}{\captionExtendedDataOne}\label{fig:extended-data-1}
\end{minipage}
\par

\clearpage
\noindent\begin{minipage}{\linewidth}
\makebox[\linewidth][c]{\includegraphics{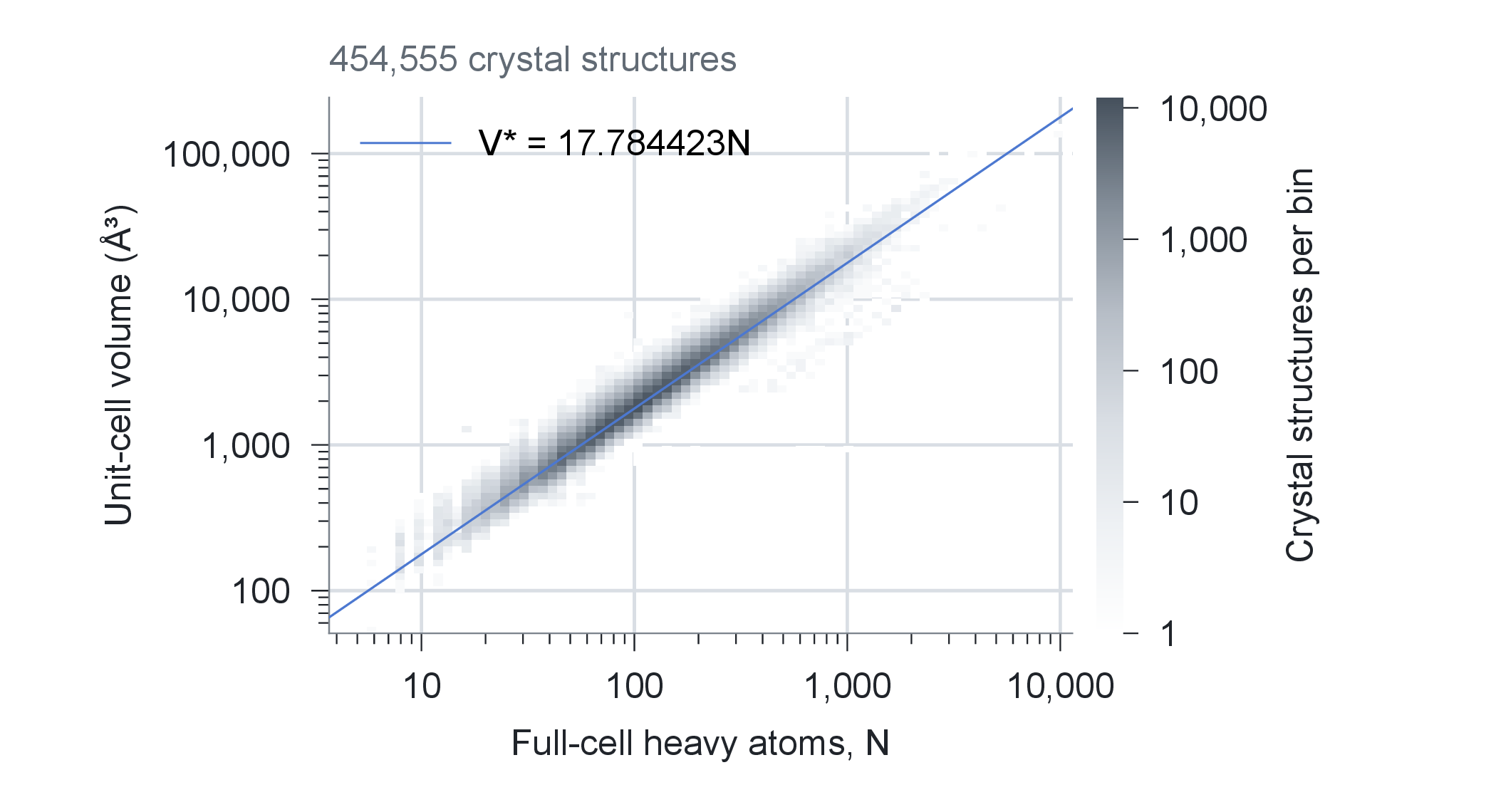}}
\captionof{figure}{\captionExtendedDataTwo}\label{fig:extended-data-2}
\end{minipage}
\par

\clearpage
\noindent\begin{minipage}{\linewidth}
\makebox[\linewidth][c]{\includegraphics{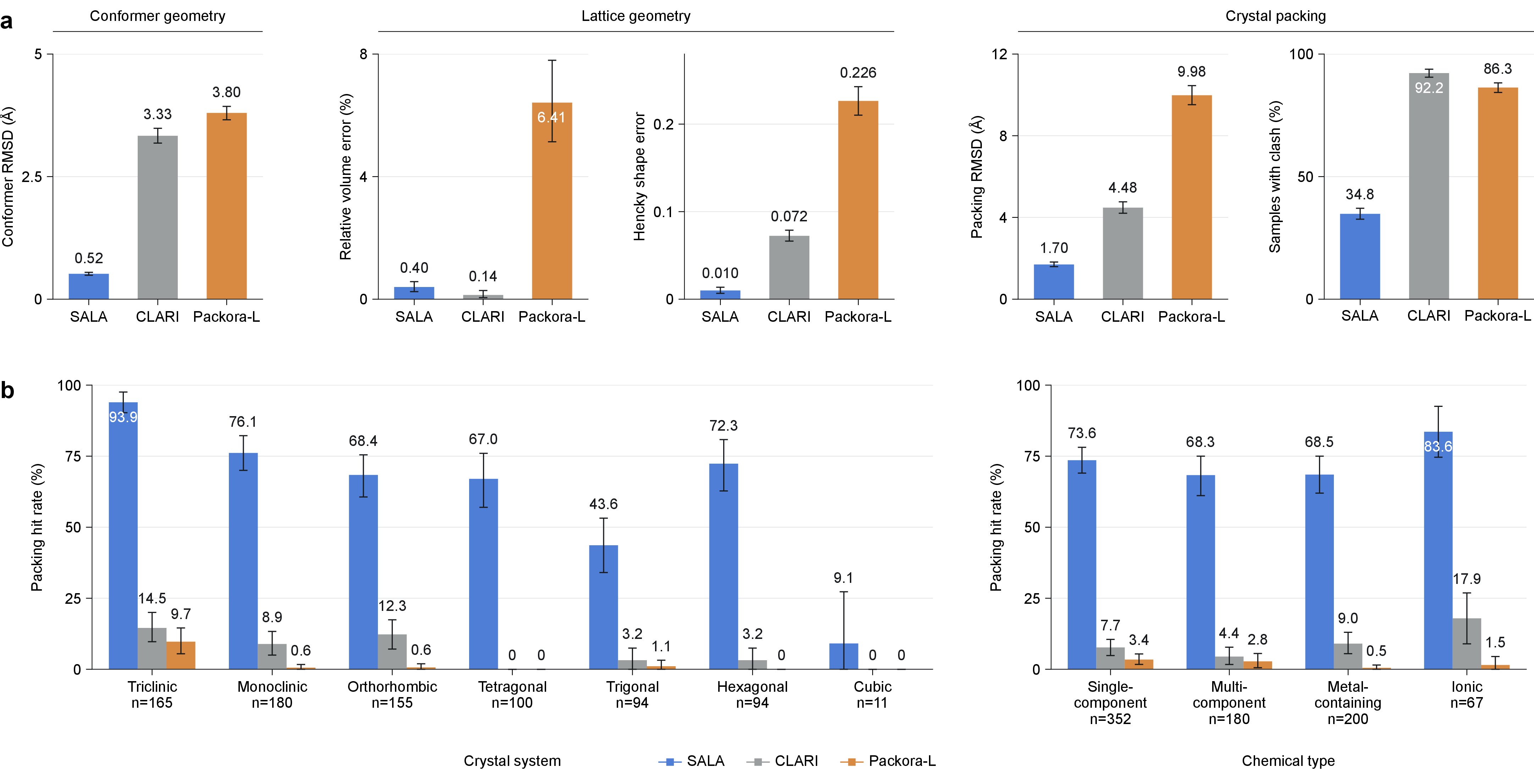}}
\captionof{figure}{\captionExtendedDataThree}\label{fig:extended-data-3}
\end{minipage}
\par

\clearpage
\noindent\begin{minipage}{\linewidth}
\makebox[\linewidth][c]{\includegraphics{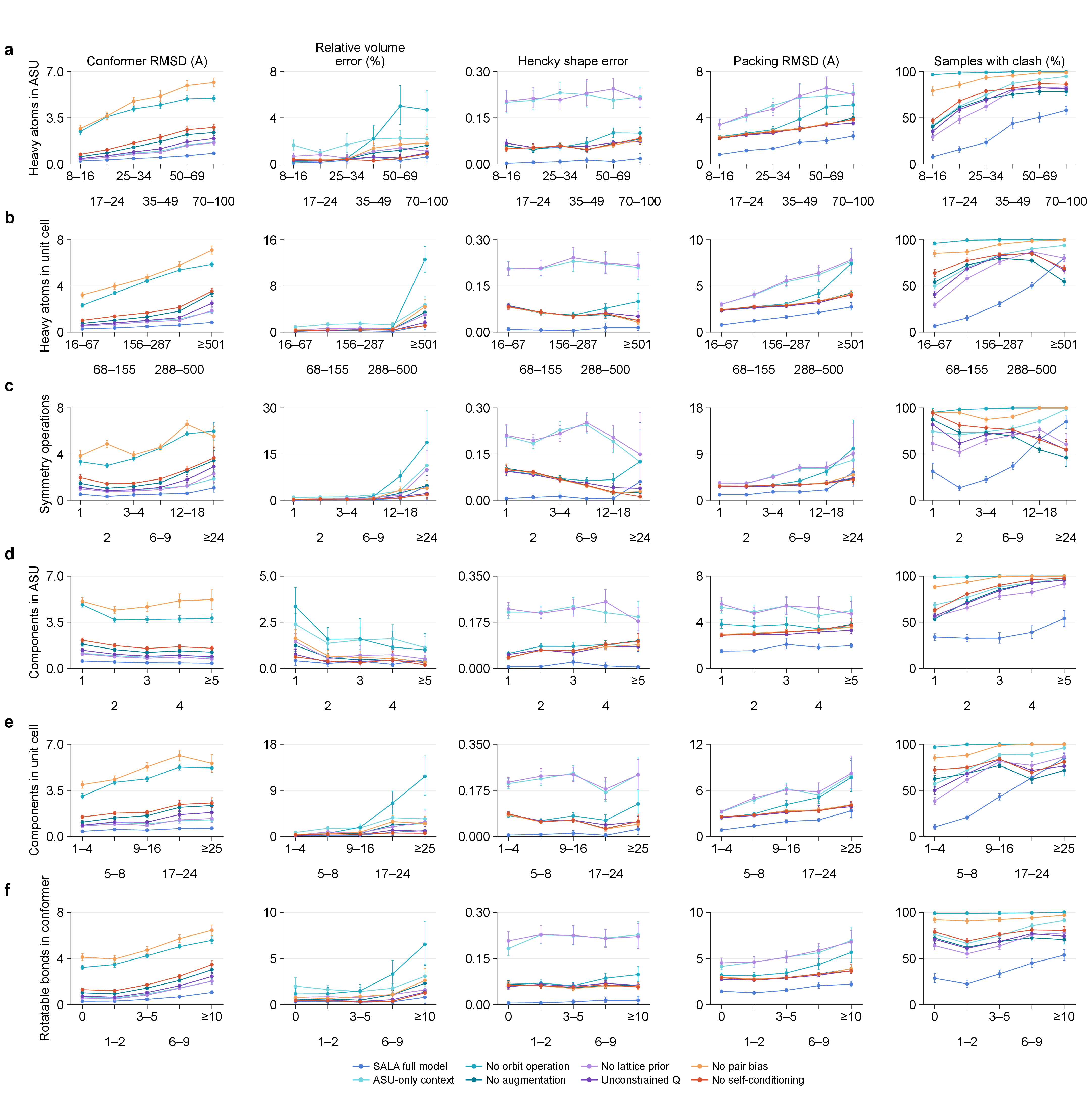}}
\captionof{figure}{\captionExtendedDataFour}\label{fig:extended-data-4}
\end{minipage}
\par
\endgroup